\documentclass[preprintnumbers,preprint,showpacs,prd]{revtex4}
\usepackage{graphicx}
\input epsf.sty

\begin{document}
\preprint{BU-HEPP-04-01}
\title{Deflated Iterative Methods for Linear Equations with Multiple
Right-Hand Sides}\thanks{This work was partially supported by the National Science Foundation, 
Computational Mathematics Program under grant 0310573 and the National Computational Science Alliance.  
It utilized the SGI Origin 2000 and IBM p690 systems at the University of Illinois.  The 
first author was also supported by the Baylor University Sabbatical Program.}

\author{Ronald B. Morgan\dag and Walter Wilcox\S}
\affiliation{\dag Department of Mathematics, Baylor
University, Waco, TX 76798-7328\\ \S Department of Physics, Baylor
University, Waco, TX 76798-7316.}

\bibliographystyle{plain}

\begin{abstract}
A new approach is discussed for solving large nonsymmetric systems of linear equations with 
multiple right-hand sides.  The first system is
solved with a deflated GMRES method that generates eigenvector information
at the same time that the linear equations are solved.  Subsequent systems
are solved by combining an iterative method with a projection over the previously determined 
eigenvectors.  Restarted GMRES is considered for the iterative method as well as non-restarted 
methods such as BiCGSTAB.  These methods offer an alternative to block methods, and they can also be
combined  with a block approach.  An example is given showing significant improvement for a problem
from quantum chromodynamics.
\end{abstract}


\maketitle

\section{Introduction}

Large systems of linear equations $Ax=b$ arise in many areas of science.  Often
there are many right-hand sides associated with a single matrix.  It
is then important to consider these systems together and take advantage of the
relationship between them.  Here we consider solving the systems with iterative methods, 
and we assume the matrix is real nonsymmetric or complex non-Hermitian.
A standard way of dealing with multiple right-hand sides is to use a
block method.  Krylov subspaces are generated with each right-hand
side as starting vector and are used together.  An alternative was presented in~\cite{SiGa}, 
with the right-hand sides solved individually using Richardson iteration with a polynomial generated from
GMRES~\cite{SaSc} applied to the first right-hand side.  Here we present another option.  Eigenvector
information generated while solving the first right-hand side is used to help solve the other right-hand
sides.  This approach can be helpful for difficult problems with small eigenvalues.

The first right-hand side is solved with a deflated GMRES method, which also generates 
approximations to eigenvectors.  These approximate eigenvectors can then be used to deflate eigenvalues
from the solution of the linear equations with the subsequent right-hand sides.  A fairly simple approach
yields a useful method.  Specifically, we alternate cycles of regular GMRES with projections over the
approximate eigenvectors.  We will also look at combining deflation with non-restarted methods such as
BiCGStab~\cite{vdV92}.  For situations when a block approach is particularly desirable, it is possible to
use a deflated block method both for the initial phase in which the eigenvectors are generated and for
solution of subsequent right-hand sides.

Section II reviews deflated GMRES methods and a projection that will be used. 

Section III gives the projected version of GMRES and gives examples comparing it to other methods.  
There are also experiments with a matrix from lattice quantum chromodynamics (QCD).  Section IV 
discusses non-restarted methods and deflated versions of BiCGStab.  A special approach for QCD problems
is developed.  Block approaches are in section V.

\section{Deflated GMRES and Projections}

Small subspaces for restarted GMRES can slow convergence for difficult problems.  
Deflated versions of restarted GMRES~\cite{GMRES-E,KhYe,ErBuPo,ChSa,Sa95B,BaCaGoRe,BuEr,
LCMo,DS99,GMRES-IR,GMRES-DR} can improve this, when the problem is difficult due to a few
small eigenvalues.  One of these approaches is related to Sorensen's
implicitly restarted Arnoldi method for eigenvalues~\cite{So} and is called
GMRES with implicit restarting~\cite{GMRES-IR}.  A mathematically
equivalent method, called GMRES with deflated restarting
(GMRES-DR)~\cite{GMRES-DR}, is also related to Wu and Simon's restarted
Arnoldi eigenvalue method~\cite{WuSi}.  See~\cite{Arnoldi-R,St01,HRAM} for
some other related eigenvalue methods.

We will concentrate on GMRES-DR, because it is efficient and relatively
simple.  Approximate eigenvectors corresponding to the small eigenvalues are
computed at the end of each cycle and are put at the beginning of the next
subspace.  Letting $r_0$ be the initial residual for the linear equations
at the start of the new cycle and $\tilde y_1, \ldots
\tilde y_k$ be harmonic Ritz vectors~\cite{IE,Fr92,PaPavdV,IEN}, the
subspace of dimension $m$ used for the new cycle of GMRES-DR(m,k) is

\begin{equation}
Span\{\tilde y_1, \tilde y_2, \ldots \tilde y_k, r_0, A r_0, A^2 r_0, A^3
r_0, \ldots ,A^{m-k-1} r_0 \}. \label{ss}
\end{equation}

This can be viewed as a Krylov subspace generated with starting vector
$r_0$ augmented with approximate eigenvectors.  Remarkably, the whole
subspace turns out to be a Krylov subspace itself (though not with $r_0$ as
starting vector)~\cite{GMRES-IR}.
Once the approximate eigenvectors are moderately accurate, their inclusion
in the subspace for GMRES essentially deflates the corresponding
eigenvalues from the linear equations problem.

The approximate eigenvectors in GMRES-DR span a small Krylov subspace and so are generated in 
a compact form 

\begin{equation}
 AV_k = V_{k+1} \bar H_k, \label{recur1}
\end{equation} 

where $V_k$ is a $n$ by $k$ matrix whose columns span the subspace of
approximate eigenvectors, $V_{k+1}$ is the same except for an extra column
and $\bar H_k$ is a full $k+1$ by $k$ matrix.   Note this compact form is
similar to an Arnoldi recurrence, and it allows access to both the
approximate eigenvectors and their products with $A$ while requiring
storage of only $k+1$ vectors of length $n$.

Next we give the specific minimum residual (minres) projection which will be needed.  
Here the projection is over the subspace spanned by the columns of the matrix $V_k$ of approximate
eigenvectors from Equation~(\ref{recur1}).  See Saad~\cite{Sa96} for more on projections.

\vspace{.10in}

\begin{center}

\textbf{Minres Projection}

\end{center}

\begin{enumerate}

 \item Let the current approximate solution be $x_0$ and the current system

of equations be $A(x-x_0) = r_0$.  Let $V_{k+1}$ be the Arnoldi-type matrix from Equation (\ref{recur1}).  

 \item Solve min$||c - \bar H_k d||$, where $c = (V_{k+1})^T r_0$.

 \item The new approximate solution is $x_k = x_0 + V_{k}d$.

 \item The new residual vector is $r_k = r_0 - AV_{k} d = r_0 - V_{k+1} \bar H_k d$.

\end{enumerate} 

\vspace{.15in}

This projection is fairly inexpensive, requiring only $3k+2$ vector operations (dot products and daxpys) 
of length $n$.

\section{Deflated GMRES for Multiple Right-hand Sides}

If a method such as GMRES-DR is used for the first right-hand side, eigenvector information is generated 
while the linear equations are solved.  We wish to use this information to assist with the solution of the
other right-hand sides.  We will suggest three ways of doing this.  The main focus will be on the third
approach, and it will be compared against the first two.

One possible way to use the approximate eigenvectors is to put them into the subspaces used for GMRES.  
Such a method is called GMRES-E in~\cite{GMRES-E}.  The subspace has a basis like~(\ref{ss}) for GMRES-DR,
but with the approximate eigenvectors going last in forming the basis. Also, since the eigenvectors are
already computed, they can be left fixed while solving the subsequent right-hand sides.

Another approach to deflating eigenvalues for the subsequent right-hand sides is to use the approximate 
eigenvectors to build a preconditioner for GMRES.  Burrage and Erhel propose a method called
DEFLATION~\cite{BuEr}.  They do not consider multiple right-hand sides, but their method can be adapted by
using the preconditioner from DEFLATION, but not the portion of DEFLATION that computes eigenvectors.  For
both these approaches to deflation, there are significant costs compared to simple restarted GMRES.  With
GMRES-E, there are $k$ additional vectors that are orthonormalized.  With DEFLATION, every iteration
requires additional work in applying the preconditioner.  The approach we discuss next is more efficient.

A relatively simple way of deflating is to use the projection mentioned in the previous section.  
Projections over the subspace of approximate eigenvectors can be alternated with cycles of GMRES.  A major
difference between this approach and  those mentioned in the last two paragraphs is that the eigenvectors
are not needed during the GMRES iteration.  This approach can be much cheaper if many eigenvectors are
used.  We call this method GMRES(m)-Proj(k), where $m$ is the dimension of the Krylov subspaces used in the
GMRES cycles and $k$ is the number of approximate eigenvectors.  These projections are mentioned briefly
in~\cite{GMRES-DR} for the case of just one right-hand side.  For multiple right-hand sides, some
preliminary experiments are reported in~\cite{Qcdconf} for lattice QCD problems.  Further QCD experiments
are at the end of this section.

The GMRES-Proj method that follows is for all right-hand sides except for the first one.  
Superscripts identify which right-hand side the vectors are associated with.  

\vspace{.10in}

\begin{center}

\textbf{GMRES(m)-Proj(k)}

\end{center}

\begin{enumerate}

 \item After applying the initial guess $x_0^{(i)}$, let the system

of equations be $A(x^{(i)}-x_0^{(i)}) = r_0^{(i)}$.  

 \item If it is known that the right-hand sides are related, project over the previous computed 
solution vectors.

 \item Apply the Minres Projection using the $V_{k+1}$ and $\bar H_k$ matrices developed while 
solving the first right-hand side with GMRES-DR.

 \item Apply one cycle of GMRES(m).

 \item Test the residual norm for convergence (can also test during the GMRES cycles).  If not 
satisfied, go back to step 3.

\end{enumerate} 

\vspace{.15in}

The projection step adds little to the cost of the method.  One cycle of GMRES requires about $m^2 + 2m$ 
length $n$ vector operations plus the cost of $m$ matrix-vector products and applications of the
preconditioner.  The projection step uses just over $3k$ vector ops and requires no matrix-vector
products.  If for example $m=15$, $k=10$, the matrix has five nonzeros per row and no preconditioning is
used, then the projection adds only $10 \%$ additional cost to a cycle.

\subsection{Experiment}

The first example uses a simple test matrix for which deflation is important, because there are 
some small eigenvalues.

{\it Example 1.}

\begin{figure}
\vspace{.10in}
\includegraphics[width=5.25in]{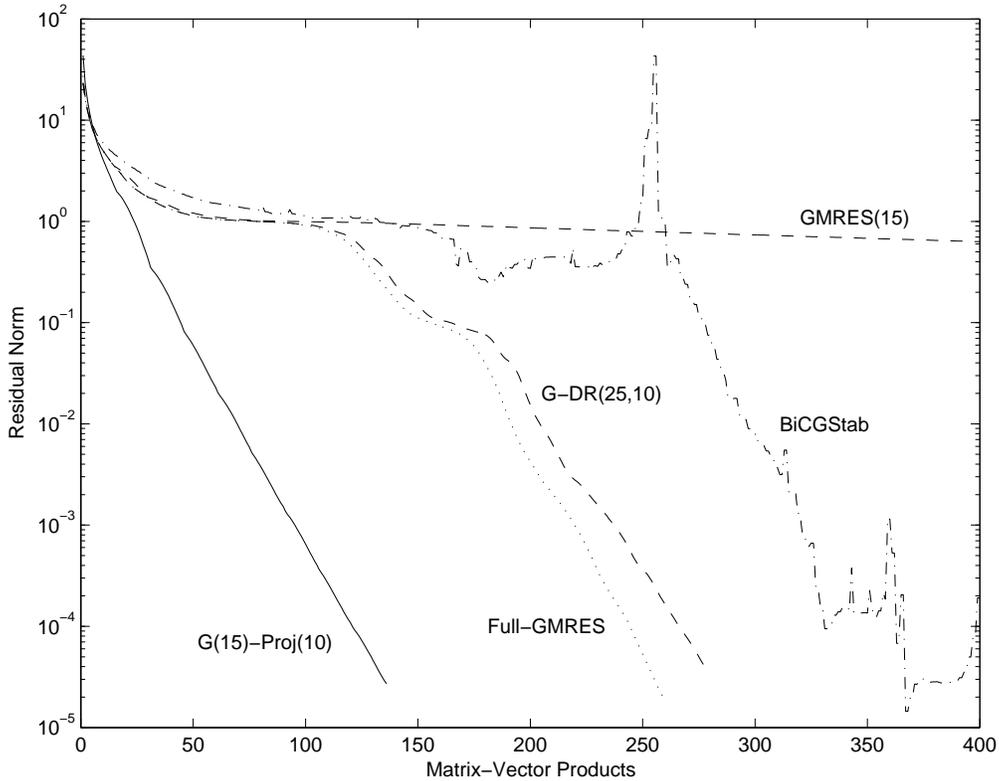}
\vspace{.10in}
\caption{Solution of second right-hand side.}
\label{one}
\end{figure}

The matrix is of size $n=2000$ and is bidiagonal with $0.1, 1, 2, 3, \ldots,$ $1998, 1999$ on the 
main diagonal and $1$'s on the superdiagonal.  GMRES-DR(25,10) has subspaces of total dimension 25
including 10 approximate eigenvectors.  It is applied to a randomly generated first right-hand side (random
with unit normal distribution) until the residual norm has improved by a factor of $rtol = 10^{-6}$.  This
takes 280 matrix-vector products. GMRES(15)-Proj(10) is then applied to a random second right-hand side. 
We compare with several other methods, BiCGStab, Full GMRES, GMRES-DR(25,10) and GMRES(15).  The results
for this second right-hand side are given in Figure \ref{one}.  Notice that GMRES-Proj has a big advantage over
the other methods, because it deflates eigenvalues from the beginning.  The methods Full-GMRES, BiCGStab,
and GMRES-DR must generate approximate eigenvectors as they proceed.  GMRES(15) restarts before it can
develop effective approximate eigenvectors.

We consider both matrix-vector products and flops, so that two cases can be simulated with this one 
test matrix.  For problems with expensive matrix-vector product or preconditioner, the matrix-vector
product count matters.  For very sparse matrices without preconditioner, the flop count for this sparse
test matrix is more relevant.  Of course many problems fall in between these extreme cases.

For the first right-hand side, BiCGStab uses more matrix-vector products than GMRES-DR but 
considerably less flops.  BiCGStab needs 17.5 million flops versus 81.1 million for GMRES-DR to improve the
residual norm by $10^{-6}$.  However, GMRES-Proj saves on both matrix-vector products and flops for the
second right-hand side.  It uses 130 matrix-vector products compared to 365 for BiCGStab and 14.1 million
flops versus 17.5.  Of course GMRES-Proj needs GMRES-DR applied first, but if there are a number of
right-hand sides, the GMRES approach can still be competitive in terms of flops even for such a very sparse
matrix.  For example, if there are 10 right-hand sides, then GMRES-DR for the first right-hand side and
GMRES-Proj for the next nine takes 1405 matrix-vector products and 204.5 million flops.  BiCGStab on all 10
right-hand sides uses 4113 matrix-vector products and 197.6 million flops.

\subsection{Effect in GMRES-Proj of the size of the GMRES subspace}

We experiment with changing the size of the GMRES subspaces used to solve the subsequent 
right-hand sides.  The same problem from the experiment in the previous subsection is considered with again
the first right-hand side solved with GMRES-DR(25,10).  This time ten right-hand sides are solved, with
different $m$ values for GMRES(m)-Proj(10).  The first column of Table \ref{changingfreq} gives the total
number of matrix-vector products required to solve all ten systems.  We see that even with deflation, larger
subspaces are helpful.  However, if the matrix-vector product is inexpensive, using a small value of $m$
such as $m=10$ might be more efficient than $m=20$ in spite of increased iterations.

\begin{table}

\caption{Changing $m$ and the frequency of projection for GMRES-Proj} 

\begin{center} \footnotesize

\begin{tabular}{|c|c|c|c|c|}  \hline

       & project every cycle   & project every 5th  & project every 10th & project at 10,20, ... \\ \hline

$m$    & mat-vec's  & mat-vec's & mat-vec's & mat-vec's \\  

\hline\hline

5      & 2440    & 2577 & 2596 & 1962 \\ \hline

10     & 1658    & 1678 & 1636 & 1604 \\ \hline

15     & 1405    & 1411 & 1673 & 1910 \\ \hline

20     & 1298    & 1330 & 2107 & 2438 \\ \hline 

25     & 1257    & 1478 & 2521 & 2843 \\ \hline

\hline

\end{tabular} 

\end{center} 

\label{changingfreq} 

\end{table}

\subsection{Projecting less frequently}

Table \ref{changingfreq} also shows the effect of not projecting between every cycle of GMRES(m). For small
values  of $m$, it is not necessary to project very often.  Projecting reduces components of $r$ in the
directions of the  eigenvectors corresponding to small eigenvalues, and these components may not need to be
reduced futher for a while (basically until the rest of the residual vector has been reduced to the point that
these components are again significant).  For $m=10$, projecting every tenth cycle is good enough.  For
$m=20$, projecting every fifth is almost as good as projecting every cycle, while projecting every tenth is
not as effective.

In the case of $m=5$ and project every tenth cycle, the projections 

are performed before the 1st, 11th, 21st, \ldots cycles of GMRES.  If instead we project before the 
10th, 20th, \ldots cycles, the results are better, with only 1962 matrix-vector products needed instead of
2596 (see the last column of Table \ref{changingfreq}).  It is not clear why the convergence is better if no
projection is performed until after nine cycles of GMRES.   

\subsection{Solving the first right-hand side to greater accuracy}

We test here the notion that the eigenvector approximations provided by GMRES-DR might not be 
optimal at the same point that the linear equations are considered solved.  For the case of Example 1,
solving 10 right-hand sides with $rtol$ of $10^{-6}$ requires 1405 iterations.  But if the system with the
first right-hand side is solved to greater accuracy of $10^{-8}$, while the final nine systems are again
are solved to $10^{-6}$, the total number of iterations drops to 1343.  This is in spite of the fact that
solution of the first system takes 69 more iterations.  The average savings per subsequent right-hand side
is 13.6 iterations.  However, solving the first right-hand side to even greater accuracy does not pay off. 
With relative tolerance of $10^{-10}$, 1388 iterations are needed for all ten right-hand sides (exempting
the first, the number actually stays the same as for the first $rtol$ being $10^{-8}$).

\subsection{Comparison with other deflation approaches}

While the GMRES-Proj method deflates eigenvalues with a projection that is separate from the GMRES phase, 
there are other ways of deflating eigenvalues as discussed at the beginning of this section.  We will
compare GMRES-Proj with the versions of GMRES-E and DEFLATION that use eigenvectors to augment or
precondition, but are adapted so they do not attempt to improve on the eigenvectors.  Note GMRES-DR is used
on the first right-hand side to compute the eigenvectors for each method, then nine additional right-hand
sides are solved.  We see from the results in Table \ref{compdef} that the methods perform similarly.  However,
as mentioned earlier, there is a difference in expense, since GMRES-Proj uses eigenvectors only once per
cycle.  DEFLATION applies eigenvectors at every iteration, and the cost above the normal GMRES expense is
$2k$ length $n$ vector operations per iteration or about $2km$ per cycle.  Costs for GMRES-E with $k$
approximate eigenvectors augmenting a $m$-dimensional Krylov subspace are a little greater than for
DEFLATION (extra expense of about $2km + k^2$ length $n$ vector operations per cycle).   Meanwhile, as
mentioned, GMRES-Proj requires about $3k$ extra per cycle.  So GMRES-Proj can be more efficient.  However,
for very expensive matrix-vector product or for small $k$, GMRES-Proj may not be a significantly better way
of deflating.

\begin{table}

\caption{Comparson of different deflation approaches} 

\begin{center} \footnotesize

\begin{tabular}{|c|c|c|c|}  \hline

       & eigenvectors used for projection: &  eigenvectors in subspace: & eigenvector preconditioner: \\ 

$m$    & GMRES-Proj  &  GMRES-E            & DEFLATION  \\  

\hline

5      & 2440    & 2565 & 2528 \\ \hline

10     & 1658    & 1658 & 1613 \\ \hline

15     & 1405    & 1405 & 1387 \\ \hline

20     & 1298    & 1296 & 1284 \\ \hline 

25     & 1257    & 1251 & 1241 \\ \hline

\hline

\end{tabular} 

\end{center} 

\label{compdef} 

\end{table} 

\subsection{Comparison with block-QMR}

Here we show that the GMRES-Proj method can be competitive with block methods.  Specifically, we 
compare against block-QMR~\cite{FrMa} for 10 right-hand sides.  GMRES(15)-Proj(10) uses projections every
fifth GMRES cycle.  Table \ref{compnon} has the results for both the number of matrix-vector products and the
number of flops as counted by MATLAB.  GMRES-Proj is a little better than block-QMR in terms of flops, since
block-QMR with blocksize 10 has significant orthogonalization expense.  (Table \ref{blockGMRES} includes
comparison with 40 right-hand sides.)

\begin{table}

\caption{Comparison with nonrestarted methods including block QMR} 

\begin{center} \footnotesize

\begin{tabular}{|c|c|c|c|}  \hline

 Method      & matrix-vector products  &  Mflops     \\  

\hline

GMRES-DR + GMRES-Proj  & 1411    & 198.3   \\ \hline

Block-QMR              & 1782    & 567.5   \\ \hline

QMR, 10 times          & 5220    & 245.7   \\ \hline

BiCGStab, 10 times     & 4113    & 197.6   \\ \hline 

\hline

\end{tabular} 

\end{center} 

\label{compnon}

\end{table}

\subsection{The case of related right-hand sides}

One expects intuitively that if the right-hand sides are closely related to each other then there 
should be a way to take advantage of the situation.  However, it is discussed in~\cite{bgdr} that block
methods may not be successful at this.  We suggest here a simple way for GMRES-Proj to deal with this
case.  For the second and subsequent right-hand sides, projections are done over all previously computed
solutions (step 2 of the GMRES-Proj algorithm).  We project over each solution vector individually, but
another option is to project over all at once.  

We again compare GMRES-Proj with block-QMR for 10 right-hand sides.  This time the first has random 
normal entries and all the others are equal to the first one plus $10^{-4}$ times a random vector. 
GMRES-Proj is better able to take advantage of the related right-hand sides, because it solves them
sequentially, and the results of one solution are available for the next problem.  GMRES-Proj uses only 521
matrix-vector products compared to 1702 for block-QMR.  In terms of flops, GMRES-Proj needs 110 million
versus to 542 million for block-QMR.

\subsection{A QCD Example}

We demonstrate the GMRES-Proj method for an application from particle physics.  In lattice quantum 
chromodynamics (QCD), very large systems of linear equations arise that have complex non-Hermitian
matrices.  For such matrices, we need to change transpose to Hermitian transpose in the algorithms.  Often
there are multiple right-hand sides for each QCD matrix.  However, block methods are not typically used. 
The matrix-vector product is moderately expensive (it can be implemented for a cost equivalent to 72
non-zeros per row~\cite{FrMe} even though there would actually be about three times as many non-zeros in
the matrix if it was formed).  The orthogonalization costs are significant enough to discourage block
methods.  Therefore it would be very useful to improve convergence of the main methods used  for QCD
problems such as restarted GMRES and BiCGStab.  Our application of deflation to multiple right-hand sides
is new; however, deflation in the context of lattice problems was originally considered in~\cite{dF}. 
See~\cite{EdHeNa,DoLeLiZh,NeEiLiNeSc} for other approaches.

{\it Example 2.}

\begin{figure}
\vspace{.10in}
\includegraphics[width=5.25in]{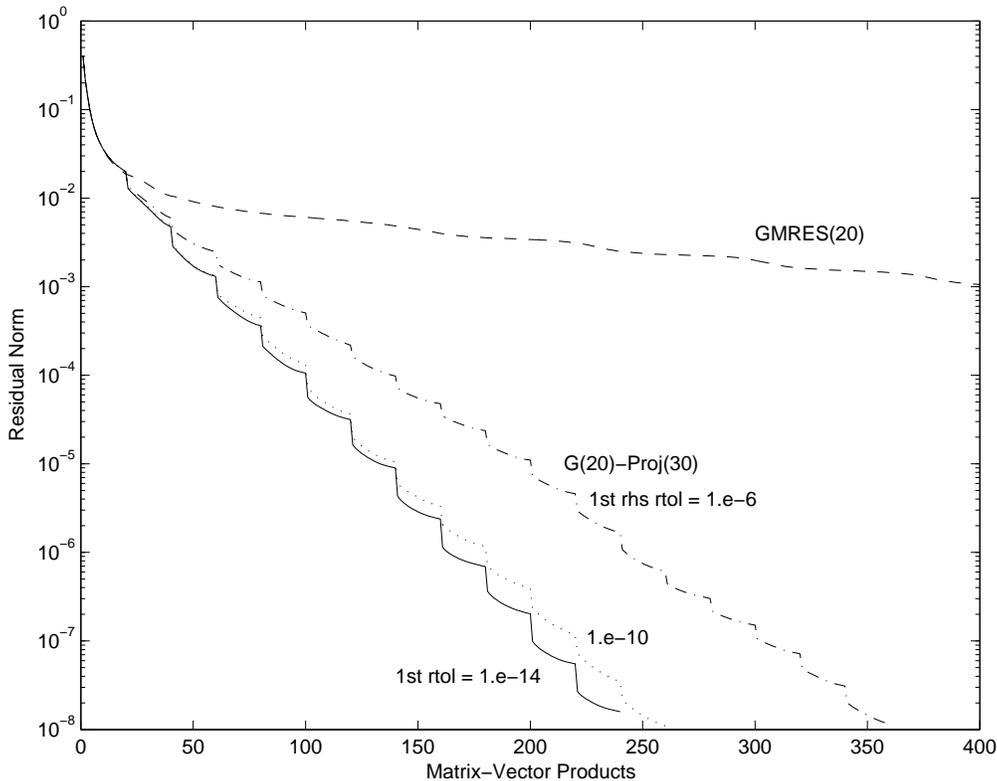}
\vspace{.10in}
\caption{Solution of second RHS for large QCD matrix.}
\label{two}
\end{figure}

We look at a typical Wilson-Dirac matrix from QCD.  It has even-odd preconditioning~\cite{Qcdconf}, 
and the dimension is 248,832 by 248,832.  The value of $\kappa$ is 0.159, which is approximately
$\kappa_{critical}$, so the leftmost eigenvalues are near the imaginary axis.  The right-hand sides are
unit vectors associated with particular space-time, Dirac and color coordinates.  The first right-hand side
is solved with GMRES-DR(50,30) to three different residual tolerances.  Then for the second right-hand
side, GMRES-Proj uses 30 approximate eigenvectors for the projection in between cycles of GMRES(20).  See
Figure \ref{two} for the results.  Solving the first right-hand side to one of the more demanding tolerances
($10^{-10}$ or $10^{-14}$) pays off.  GMRES(20)-Proj(30) can converge in less than one-tenth of the
iterations needed for GMRES(20).

\begin{figure}
\vspace{.10in}
\includegraphics[width=5.25in]{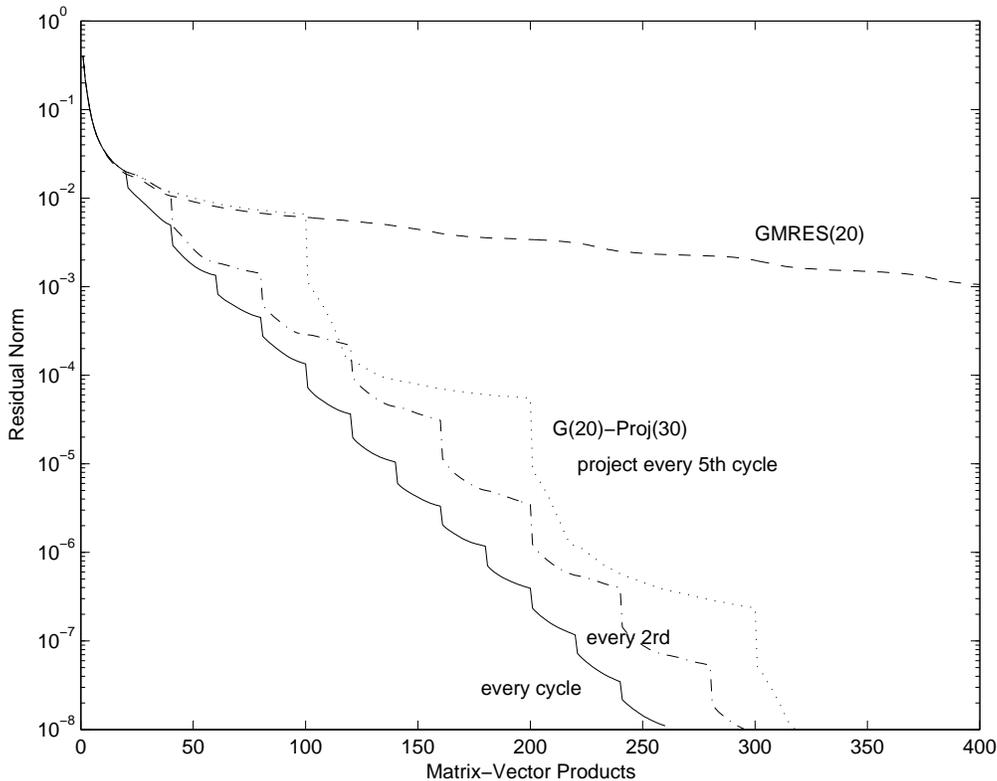}
\vspace{.10in}
\caption{Different frequencies of projection for the QCD matrix.}
\label{three}
\end{figure}

Figure \ref{three} shows convergence with different frequencies of projection for GMRES(20)-Proj(30) with the 
first right-hand side solved to $rtol = 10^{-10}$ .  Projecting in between every cycle turns out a little
better (for a different QCD matrix in~\cite{Qcdconf}, projecting every third cycle was as effective as
every cycle).

\begin{figure}
\vspace{.10in}
\includegraphics[width=5.25in]{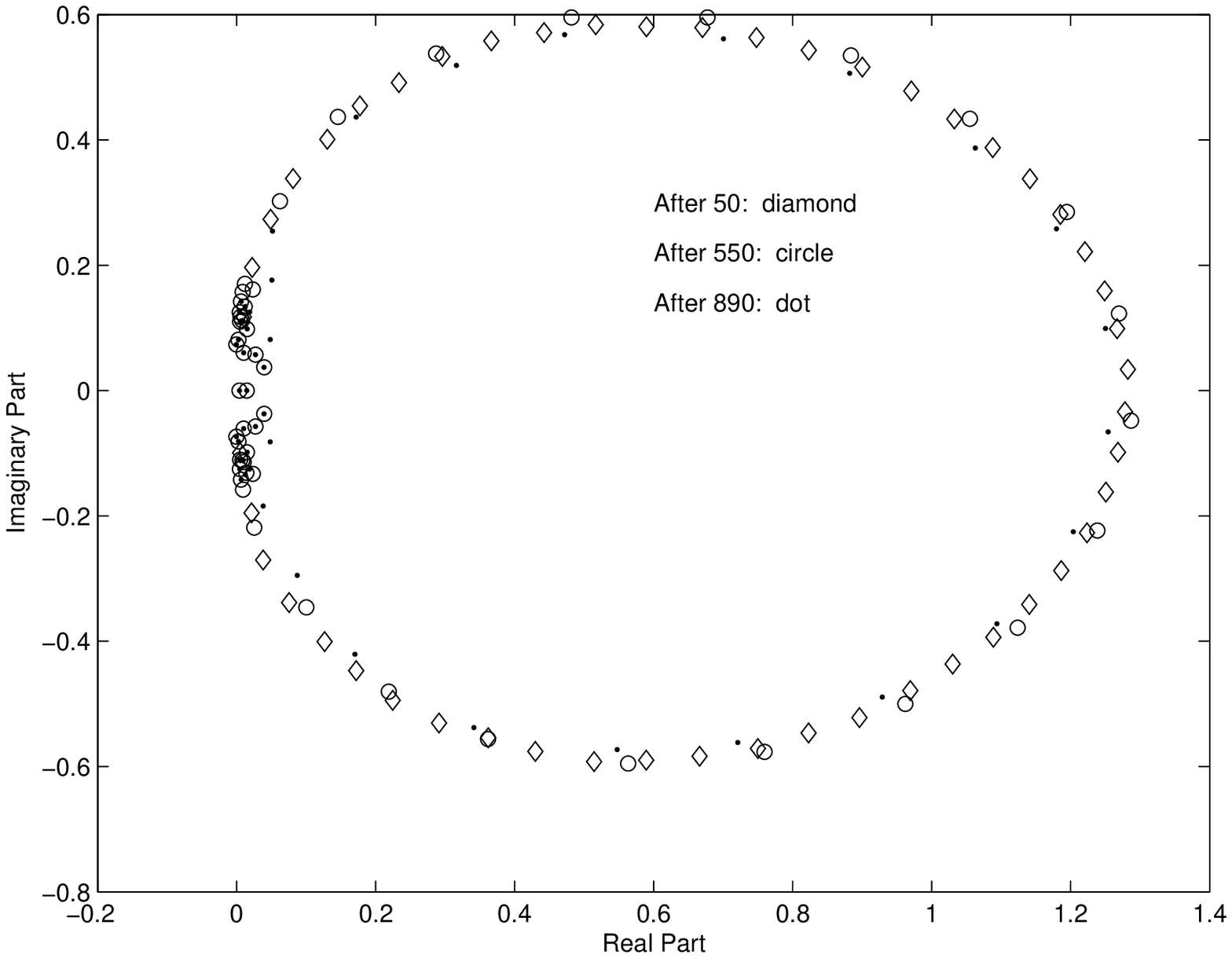}
\vspace{.10in}
\caption{Harmonic Ritz values from GMRES-DR with the QCD matrix.}
\label{four}
\end{figure}

\begin{figure}
\vspace{.10in}
\includegraphics[width=5.25in]{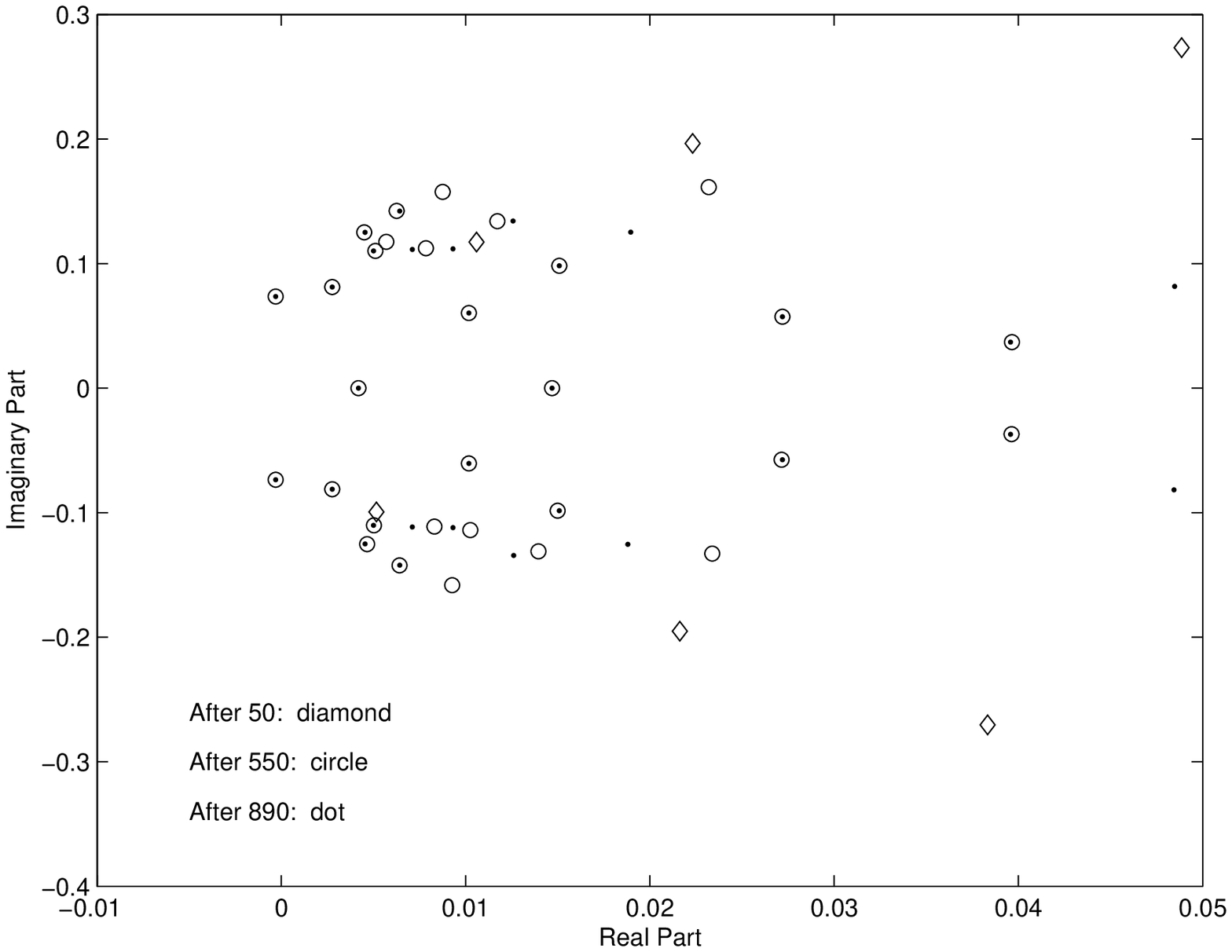}
\vspace{.10in}
\caption{Closeup of harmonic Ritz values near the origin.}
\label{five}
\end{figure}

Figure \ref{four} shows the harmonic Ritz values generated by GMRES-DR(50,30) after 50, 550 and 890 
matrix-vector products.  The case of 550 corresponds to $rtol = 10^{-10}$.  Figure \ref{five} has a blowup of
the portion of that graph near the origin.  After 550 iterations, the small approximate eigenvalues are
settling in near to where they are after 890.  These graphs show why deflating eigenvalues is so effective
for this problem.  The origin is halfway surrounded by eigenvalues, until the smallest ones are deflated.

\section{Deflation for Nonrestarted Methods}

\subsection{A simple deflated BiCGSTAB}

Nonrestarted methods are popular, because they use a large Krylov subspace but don't have 
excessive expense or storage.  For difficult problems, nonrestarted methods often converge much faster than
regular restarted GMRES.  Deflated versions of GMRES are usually more competitive, but there are still
problems for which nonrestarted methods are preferable.  This is particularly true for the case of fairly
inexpensive matrix-vector product, or when storage is limited.  We now look at using deflation to improve
BiCGSTAB and other nonrestarted methods for multiple right-hand sides.  As in the previous section, we plan
to apply GMRES-DR to the first right-hand side and then use the eigenvector information generated to assist
with the other right-hand sides.  However, a projection is only possible at a restart.  Therefore a
nonrestarted iterative method must rely on just one projection, before the iteration begins.  We note that
deflation could be applied during the iteration with an eigenvector preconditioner~\cite{BuEr} and possibly
similar to~\cite{SaYeErGu}, where eigenvectors are worked into the conjugate gradient algorithm. 
However, here we only consider using a projection over the subspace of approximate eigenvectors.  A
deflated version of BiCGStab that can be applied to the second and subsequent right-hand sides is now
given.  It will later be modified.

\vspace{.10in}
\begin{center}
\textbf{Preliminary BiCGStab-Proj(k)}
\end{center}

\begin{enumerate}

 \item After applying the initial guess $x_0^{(i)}$, let the system

of equations be $A(x^{(i)}-x_0^{(i)}) = r_0^{(i)}$.  

 \item Apply the minres projection using the $V_{k+1}$ and $\bar H_k$ matrices developed while 
solving the first right-hand side with GMRES-DR.

 \item Apply BiCGStab until satisfied with convergence.

 \end{enumerate} 

\vspace{.15in}

\begin{figure}
\vspace{.10in}
\includegraphics[width=5.25in]{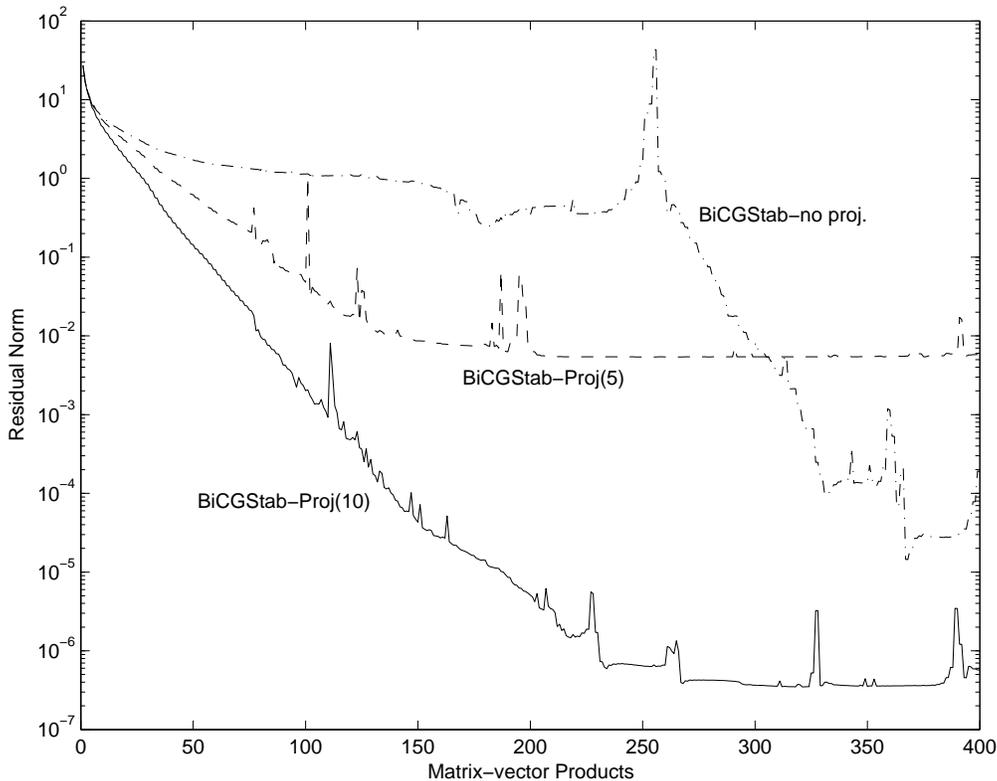}
\vspace{.10in}
\caption{Deflation for BiCGStab using only right eigenvectors.}
\label{six}
\end{figure}

{\it Example 3.}  We use the same matrix and right-hand sides as in Example 1.  The first 
right-hand side is solved using GMRES-DR with $rtol = 10^{-8}$.  Ten eigenvectors are calculated with
GMRES-DR(25,10) and five with GMRES-DR(25,5).  Figure \ref{six} shows the effect of projecting before
BiCGSTAB.  With projection of 10 eigenvectors, the method is effective.  With five eigenvectors, the projection
actually makes things worse.  The good effect appears to wear off.  This is because the components of the
residual vector in the directions of the the eigenvectors corresponding to the smallest eigenvalues are not
reduced enough by the projection.  The BiCGStab iteration must eventually confront the small
eigencomponents.  Partially removing these components can make it more difficult for BiCGStab to develop
these eigenvectors, since then BiCGStab has a poor starting vector for that task.  So partially removing
eigencomponents can slow down eventual convergence.

The minres projection simply does not do a good job.  Even though in some sense it is the best 
projection, it does not necessarily reduce small eigencomponents enough.  To see the problem with the
minres projection, we assume that we have one exact real eigenvector.  We would like to be able to
eliminate the component of the residual vector in that direction, but the minres projection does not
necessarily accomplish this.  

{\em THEOREM 4.1}.  Suppose $A$ has a full set of eigenvectors.  Let $z_1, z_2, \ldots, z_n$ be the
normalized  eigenvectors.  Let the current linear equations problem be $A(x-x_0) = r_0$, with $r_0 = 
\alpha_1 z_1 +
\alpha_2 z_2 + \ldots + \alpha_n z_n$.  Then after the minres projection over the subspace 
$Span\{z_1\}$,
the new residual's component in the direction of $z_1$ is 
\begin{equation}
-\sum_{i=2}^n \alpha_i z_1^T z_i \label{z1comp}.
\end{equation}

{\em Proof}.  Let $V$ be an othonormal matrix with columns spanning the desired subspace (the columns
may be complex). The minres projection is equivalent~\cite{Sa96} to solving 
\begin{equation}
V^T A^T A V d = V^T A^T r_0  \label{minrespr}
\end{equation}
for $d$.  The new residual vector is then $r = r_0 - A V d$.

With projection over one exact eigenvector $z_1$, Equation (\ref{minrespr}) becomes 

\[z_1^T A^T A z_1 d = z_1^T A^T r_0,\]

which gives

\[\lambda_1^2 d = \lambda_1 \sum_{i=1}^n \alpha_i z_1^T z_i.\] 

Then after solving for $d$, and with $r = r_0 - d \lambda_1 z_1$, the $z_1$ component of $r$ is 
as given in Equation (\ref{z1comp}).

Equation \ref{z1comp} shows that the minres projection over one exact eigenvector works best if the 
components of the residual in the directions other than $z_1$ are small and if the eigenvectors are nearly
orthogonal.  For nonsymmetric matrices, this projection may only reduce the size of the $z_1$ component to
roughly the size of the other components.   There is no reason to expect the $z_1$ component to be reduced
far enough so that it will not need any further reduction from the BiCGStab iteration.

\subsection{Restarting deflated BiCGSTAB}

One way to remedy the imperfection of the minres projection is to do another projection along the 
way to further reduce the size of the components of the residual vector in the directions of the smallest
eigenvectors.  However, this requires a restart.

{\it Example 4.}  

\begin{figure}
\vspace{.10in}
\includegraphics[width=5.25in]{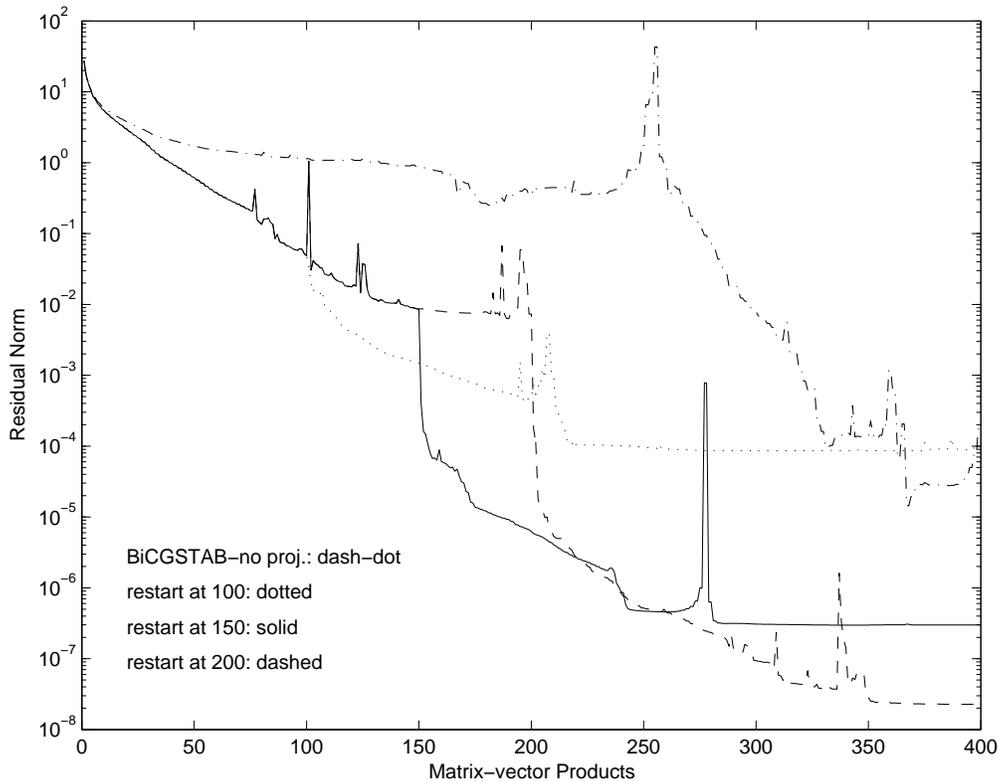}
\vspace{.10in}
\caption{Restarting deflated BiCGSTAB at different points.}
\label{seven}
\end{figure}

Figure \ref{seven} shows what happens if deflated BiCGSTAB using five eigenvectors is restarted once with an 
additional projection at the restart.  Three different runs are shown, with the restarts at 100, 150 and
200 matrix-vector products.  This approach seems fairly effective, although the restart at 100 is too early
and the residual norm plateaus around $10^{-4}$.  There may be cases where an occasionally restarted
version of deflated BiCGSTAB is useful.  The restarting would generally not need to be as frequent as for
deflated GMRES, because there are no growing orthogonalization costs for BiCGSTAB.  However, the idea of
restarting a normally nonrestarted method such as BiCGSTAB is not appealing and might slow down convergence
for difficult problems.  Also, it may be difficult to know the appropriate point or points at which to
restart.  We will next look at another possible approach that avoids restarting but needs left eigenvectors.

\subsection{A projection with left and right eigenvectors}

We give a projection over both left and right eigenvectors that does a much better job of reducing 
the eigencomponents of the residual in the directions of already determined eigenvectors.  It is simply a
Petrov-Galerkin projection~\cite{Sa96} with the right space spanned by approximate right eigenvectors and
the left space spanned by approximate left eigenvectors.

\vspace{.10in}

\begin{center}
\textbf{Left-right Projection}
\end{center}
\begin{enumerate}

 \item Let the current approximate solution be $x_0$ and the current system
of equations be $A(x-x_0) = r_0$.  Let $V$ have columns spanned by approximate right eigenvectors 
and let the columns of $W$ be spanned by approximate left eigenvectors (we choose both $V$ and $W$ to be
orthonormal).

 \item Solve $W^T A V d  = W^T r_0$.

 \item The new approximate solution is $x_k = x_0 + Vd$.

 \item The new residual vector is $r_k = r_0 - AV d $.

\end{enumerate} 

\vspace{.15in}

{\em THEOREM 4.2}.  Suppose $A$ has a full set of eigenvectors.  Let $z_1, z_2, \ldots, z_n$ be the
right eigenvectors of  length one and $u_1, u_2, \ldots, u_n$ be the left eigenvectors normalized so
that $u_i^T z_i = 1$.  Let the current linear equations problem be $A(x-x_0) = r_0$, with $r_0 =
\alpha_1 z_1 + \alpha_2 z_2 + \ldots + \alpha_n z_n$.  Then after the left-right projection with right
subspace $Span\{z_1\}$ and left subspace
$Span\{w\}$, where $w = \sum_{i=1}^n \beta_i u_i$, the new residual's component in the direction of $z_1$
is 

\begin{equation}
-\sum_{i=2}^n \alpha_i {\beta_i \over \beta_1} \label{z1comp2}.
\end{equation}

{\em Proof}.  Putting $V = z_1$ and $W = w$ into $W^T A V d = W^T r_0$, we get

\[w^T A z_1 d = w^T r_0,\]

which using the biorthogonality of the left and right eigenvectors gives

\[\lambda_1 \beta_1 d = \sum_{i=1}^n \alpha_i \beta_i.\] 

Then $d = {\alpha_1 \over \lambda_1} + {1 \over \lambda_1} \sum_{i=2}^n \alpha_i {\beta_i 
\over \beta_1}$, and with $r = r_0 - d \lambda_1 z_1$, the $z_1$ component of $r$ is as given in Equation
(\ref{z1comp2}).

This theorem tells us that the left-right projection does a better job of reducing the size of 
the residual's component in the $z_1$ direction if the other components are small and if the left vector
$w$ is mostly in the direction of the left eigenvector $u_1$.  If $w = u_1$, then the component in the
direction of $z_1$ is zeroed out.  This holds true even for projections over more than one vector.

{\em THEOREM 4.3}.  Suppose $A$ has a full set of eigenvectors.  Then with a Petrov-Galerkin projection
with $z_1$  contained in the right subspace and the corresponding left eigenvector $u_1$ in the left
subspace, the component of the residual in the direction of $z_1$ is zeroed out.

{\em Proof}.  Let the current linear equations problem be $A(x-x_0) = r_0$, with $r_0 = 
\alpha_1 z_1 + \alpha_2 z_2 + \ldots + \alpha_n z_n$.

Let $Z$ and $U$ be matrices with right and left eigenvectors respectively as columns, ordered so 
that the desired eigenvector is first and normalized so that $U^H Z = I$.

For the projection, without loss of generality $W$ can be chosen to have first column equal to 
${u_1 \over ||u_1||}.$  Let $k$ be the number of columns in $W$.  Then $W = UB$, with $B$ an $n$ by $k$
matrix with first column equal to $[{1 \over  ||u_1||}, 0, 0, \ldots 0]^T$.  Also using $A = Z \Lambda
U^H$, the problem $W^T A V d  = W^T r_0$ can be written as 

\[B^H U^H Z \Lambda U^H V d = W^T r_0.\]  

Since $\hat x = V d$, this becomes

\[B^H  \Lambda U^H \hat x = W^T r_0.\] 

Now $W^T r_0$ has first entry ${\alpha_1 \over ||u_1||}$.  Using the forms of $B$ and $\Lambda$, we 
see the solution is such that $U^H \hat x$ has first entry equal to ${\alpha_1 \over \lambda_1}$, and that
is the component of $\hat x$ in the direction of $z_1$.  The component of $A \hat x$ in the direction of
$z_1$ is then just $\alpha_1$.  So the component of $r = r_0 - A \hat x$ in that direction is zero.

\begin{figure}
\vspace{.10in}
\includegraphics[width=5.25in]{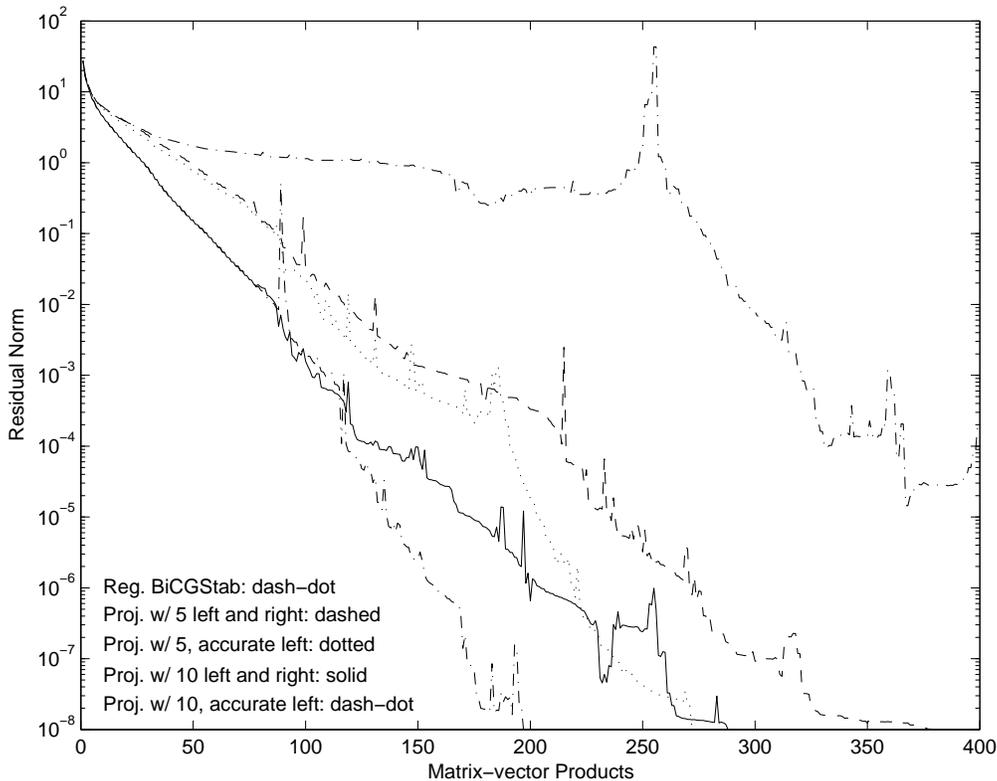}
\vspace{.10in}
\caption{Deflation for BiCGStab using right and left eigenvectors.}
\label{eight}
\end{figure}

{\it Example 5.}  We use the same matrix and right-hand sides as in Example 1.  The first right-hand 
side is solved using GMRES-DR with $rtol = 10^{-8}$.  Ten eigenvectors are calculated with GMRES-DR(25,10)
and five with GMRES-DR(25,5).  The left eigenvectors were computed separately using the same algorithms
applied to the transpose of $A$.  Figure \ref{eight} shows the effect of using the left-right projection before
BiCGSTAB.  Projecting with either five or 10 right and left approximate eigenvectors improves the result
considerably over non-deflated BiCGSTAB and also improves compared to just projecting with right
eigenvectors.  We also compute the left eigenvectors to greater accuracy (using 20 cycles of GMRES(40,20)
applied to $A^T$, then taking the eigenvectors corresponding to the smallest five or 10 eigenvalues), and
the results are even better.  On the graph these results are noted as ``accurate left".  As suggested by
Theorem 4.2, the accuracy of the left eigenvectors in the projection is important.

\subsection{The special case of QCD matrices}

For the Wilson-Dirac matrix from lattice QCD, there is a relationship between the right and left 
eigenvectors that can be exploited.  Using that a certain matrix $\gamma_5$~\cite{FrMe,Frommer} symmetrizes
the matrix $A$, it can be shown that the left eigenvector corresponding to an eigenvalue is $\gamma_5$
times the right eigenvector corresponding to the complex conjugate of the eigenvalue.  So if we calculate
the right eigenvectors for the smallest eigenvalues (including both of any complex conjugate pairs), then
$\gamma_5$ times this set gives the left eigenvectors.  Because of the simple form for $\gamma_5$, there is
no additional cost for the left eigenvectors.  We next give an example showing that deflating eigenvalues
from BiCGStab can be helpful in QCD.

{\it Example 6.}  

\begin{figure}
\vspace{.10in}
\includegraphics[width=5.25in]{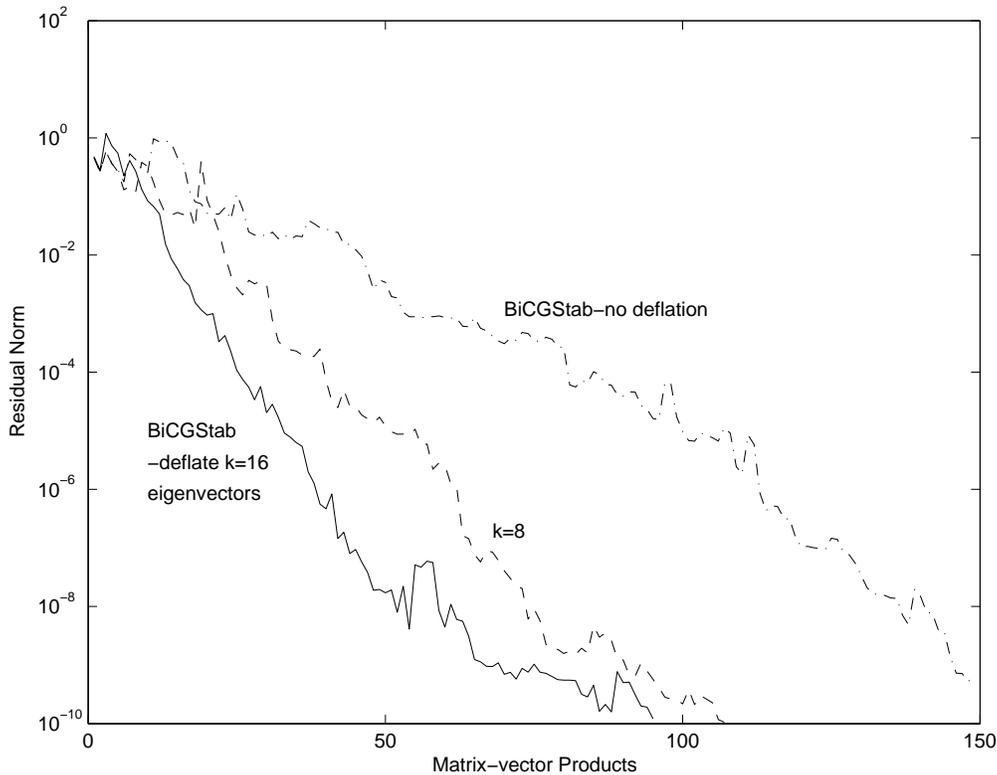}
\vspace{.10in}
\caption{Deflated BiCGStab for a small QCD example.}
\label{nine}
\end{figure}

We use a small QCD matrix.  It is complex with dimension 1536.  Solution of the second right-hand 
side has a left-right projection to deflate either eight or 16 eigenvalues followed by BiCGStab.  The right
eigenvectors are from solution of the first right-hand side with GMRES-DR and the left eigenvectors come
from the relationship with the right eigenvectors.  Figure \ref{nine} shows that to reach residual norm of
$10^{-6}$ requires only about one-third as many iterations of deflated BiCGStab as plain BiCGStab if 16
approximate eigenvectors are projected.

\section{Block Methods}

Block methods are well-known for solving systems of equations with multiple right-hand sides.  We 
saw in Subsection III.F an example for which GMRES-Proj is better than block-QMR in terms of both
matrix-vector products and flops.  However, with more right-hand sides in the next example, block-QMR is
best in terms of matrix-vector products.  So there are situations with expensive matrix-vector product
where block methods are needed.  This is particularly the case when the matrix-vector product can be
efficiently applied to several right-hand sides simultaneously.  In this section we look at combining
GMRES-Proj with block methods.

A block GMRES method with deflated restarting called block-GMRES-DR is proposed in~\cite{bgdr}.  Here 
we look at the situation where a block method is worth considering, but there are more right-hand sides
than can be efficiently solved with a single block run.  This could be either because the orthogonalization
expense or storage would be too great or because not all right-hand sides are available at once.  We
propose to solve the first group of say $p$ right-hand sides with block-GMRES-DR(m,p,k) (subspaces of
dimension $m$ are used, the block-size is $p$, and $k$ approximate eigenvectors are generated).  The
eigenvectors satisfy a block Arnoldi-like recurrence of the form $AV_k = V_{k+p}\bar H_m$, where $V_{k}$ is
an orthonormal matrix with columns spanning the space of approximate eigenvectors, $V_{k+p}$ has $p$
columns appended to $V_k$, and $\bar H_k$ is $k+p$ by $k$.  For the next group of $p$ right-hand sides, we
alternate minres projections over the approximate eigenvectors with cycles of block-GMRES.  Other
right-hand sides are dealt with the same way, $p$ at a time.

\vspace{.10in}

\begin{center}

\textbf{Bl-GMRES(m,p)-Proj(k)}

\end{center}

\begin{enumerate}

 \item Apply initial guesses to the current $p$ right-hand sides being considered. 

 \item Apply the Minres Projection to all $p$ systems using the $V_{k+p}$ and $\bar H_k$ matrices 
developed while solving the first $p$ right-hand sides with Bl-GMRES-DR(m,p,k).

 \item Apply one cycle of Bl-GMRES(m,p).

 \item Test the residual norms for convergence (can also test during the Bl-GMRES cycles).  If not 
satisfied, go back to step 2.

\end{enumerate} 

\vspace{.15in}

{\it Example 7.}  We test the matrix of Example 1 for 40 right-hand sides.  See Table \ref{blockGMRES} for the 
results.  The method Bl-G-DR(170,20,10) + Bl-G(160,20)-Proj(10) means that Block-GMRES-DR with block-size
of 20, 10 approximate eigenvectors, and total subspaces of dimension 170 is applied to the first 20
right-hand sides.  Then for the next 20 right-hand sides, minres projection over the 10 approximate
eigenvectors is alternated with Block-GMRES using block-size of 20 and subspaces of maximum dimension 160. 
However, for this problem Block-GMRES-DR does not converge.  The dimension of the Krylov subspace generated
for each right-hand side is just 8, which is not enough for this difficult problem.  With block-size of 5,
the method Bl-G-DR(170,5,10) + Bl-G(160,5)-Proj(10) uses many more flops than the non-block GMRES-Proj
approach, but it does use less matrix-vector products.  Block-QMR uses even fewer matrix-vector products,
and it can potentially take advantage of applying 40 matrix-vector products simultaneously.  Block-QMR
builds a very large subspace that eventually contains approximations to many eigenvectors, thus giving this
rapid convergence.  If one is only interested exclusively in the number of matrix-vector products,
Bl-GMRES-DR can actually be the winner.  Only 2400 matrix-vector products are needed for
Bl-GMRES-DR(1220,40,20).

\begin{table}

\caption{Comparison of block-GMRES-Proj with other methods for 40 right-hand sides}

\begin{center} \footnotesize

\begin{tabular}{|c|c|c|c|}  \hline

  Method     & matrix-vector products  &  Mflops     \\  

\hline

GMRES-DR(25,10) + GMRES(15)-Proj(10)       &  5151   &   6.1   \\ \hline

Bl-G-DR(170,20,10) + Bl-G(160,20)-Proj(10)   &   -     &  -      \\ \hline

Bl-G-DR(170,10,10) + Bl-G(160,10)-Proj(10)   &  4960   &  170.4  \\ \hline

Bl-G-DR(170,5,10) + Bl-G(160,5)-Proj(10)     &  4169   &  116.6  \\ \hline

Bl-G-DR(60,5,10) + Bl-G(50,5)-Proj(10)       &  5900   &   30.9  \\ \hline

Bl-QMR                                     &  3298   &  117.2  \\ \hline

\hline

\end{tabular} 

\end{center} 

\label{blockGMRES} 

\end{table}

\subsection{Deflating Block-QMR}

Deflation can be used to help block-QMR, as in the section on deflated BiCGStab.  We can 
use both left and right eigenvectors to remove small eigenvalue components then apply the block method. 
While improving block-QMR is an interesting idea, it is not so easy to do.

We continue the experiments in Example 7.  We use 10 right and left eigenvectors (with the accurate 
left eigenvectors from Example 5) to deflate, then apply block-QMR.  Table \ref{testdef} has the number of
matrix-vector products that are required.  With three right-hand sides, the number of matrix-vector
products is reduced significantly, from 922 to 538.  We also tried using 20 right and left eigenvectors and
the number of matrix-vector products went down to 412.  However, the improvement is not so good for larger
blocks.  Regular block-QMR manages its own deflation of eigenvalues as it builds large subspaces.

Deflated block-QMR does converge quicker at the initial stages.  For instance, if the convergence 
tolerance is only $10^{-4}$ for 10 right-hand sides, then the deflated approach reduces the number of
matrix-vector products by 31\% (from 1570 to 1076) compared to 22\% (from 1782 to 1384) for tolerance of
$10^{-6}$.

\begin{table}

\caption{Test of deflated block-QMR}

\begin{center} \footnotesize

\begin{tabular}{|c|c|c|c|}  \hline

number of right-hand sides  & Bl-QMR  &  deflated Bl-QMR      \\  

\hline

1      & 534     & 180    \\ \hline

3      & 922     & 538    \\ \hline

5      & 1256    & 838    \\ \hline

10     & 1782    & 1384   \\ \hline 

20     & 2452    & 2116   \\ \hline

40     & 3298    & 3074   \\ \hline

\hline

\end{tabular} 

\end{center} 

\label{testdef} 

\end{table}

\section{Conclusion}

In this paper, we have shown that deflating eigenvalues can be very helpful for solving systems with 
multiple right-hand sides.  The first right-hand side is solved with the deflated GMRES method GMRES-DR. 
This method develops eigenvector information that is used for all subsequent right-hand sides.  Therefore
there is no requirement that the right-hand sides all be available simultaneously.  Also, since the needed
eigenvector information is available from the beginning for the subsequent right-hand sides, the
convergence can be much faster, particularly for tough problems with small eigenvalues.  The approach in
GMRES-Proj of projecting in between cycles of GMRES is very efficient.  For the case of related right-hand
sides, there is a simple, but especially effective approach.

In lattice QCD physics, very large systems with multiple right-hand sides need to be solved.  
In one example, GMRES-Proj is an order of magnitude better that regular restarted GMRES.

Block methods are a competing approach.  However, block methods can be combined with deflation of 
eigenvalues by not solving all systems at once.

For non-restarted methods such as BiCGStab, deflating eigenvalues can also be useful.  Approximate 
eigenvectors from solving the first right-hand side with GMRES-DR can be projected one time, but both right
and left eigenvectors are needed.  For many QCD matrices, left eigenvectors are available once the right
eigenvectors have been computed.  Since BiCGStab is a popular method in QCD, this new approach should be
useful.

Future research will focus on deflating eigenvalues for QCD problems which not only have multiple 
right-hand sides, but have multiple shifts of the matrix for each right-hand side.  The goal is to solve
all the shifted systems for approximately the same cost as solving one.  It would also be worthwhile to
investigate deflation for other QCD problems such as twisted mass and overlap fermions.

For problems that are not from QCD, the need for both left and right eigenvectors for deflated BiCGStab 
makes development of a deflated restarted BiCG or QMR method desirable.  This method would solve linear
equations and simultaneously compute left and right eigenvectors.

\section*{Acknowledgments} The first author wishes to thank Andreas Frommer for helpful discussions.

\end{document}